\documentclass[12pt]{article}
\usepackage{amsmath, amsthm, amsfonts, bm}
\usepackage{amssymb}
\usepackage{mathtools}
\usepackage{float}

\usepackage{hyperref}
\hypersetup{colorlinks=true, citecolor=blue, linkcolor=blue, filecolor=blue}
\usepackage{graphicx,psfrag,epsf, float}
\usepackage{enumerate,titlesec, color}
\usepackage{natbib}
\usepackage{footnote}
\usepackage{url} 
\usepackage{soul}
\usepackage{tikz}
\usepackage{algorithm, algorithmicx, algpseudocode}

\usepackage[font=small,labelfont=bf]{caption}
\usepackage{subcaption}

\usepackage{algorithm, algorithmicx, algpseudocode}
\usepackage{booktabs}
\usepackage{lineno}

\usepackage{soul}

\usepackage{setspace}

\usepackage[textwidth=6.5in,bottom=1.8in,top=1in]{geometry}

\renewcommand\footnotemark{}

\usepackage{booktabs}

\usepackage{array}

\usepackage{multirow}

\def\bfgamma{\boldsymbol \gamma}

\begin{document}
	\bigskip
	\date{}

\title{Bayesian subtyping for multi-state brain functional connectome with application on adolescent brain cognition}
\author{Tianqi Chen$^{1}$, Chichun Tan$^{2}$, Hongyu Zhao$^1$, Todd Constable$^3$,\\
Sarah Yip$^4$, Yize Zhao$^1$\\
\small $^{1}$Department of Biostatistics, Yale University\\
\small $^{2}$Department of Biostatistics, Brown University\\
\small $^{3}$Department of Radiology and Biomedical Imaging, Yale University\\
\small $^{4}$Department of Psychiatry, Yale University
}
\maketitle

\def\spacingset#1{\renewcommand{\baselinestretch}%
	{#1}\small\normalsize} \spacingset{1}

\begin{abstract}
Converging evidence indicates that the heterogeneity of cognitive profiles  may arise through detectable alternations in brain functions. Particularly, brain functional connectivity, measured under resting and cognitive states, characterizes the unique neuronal interconnections across large-scale brain networks. Despite an unprecedented opportunity to uncover neurobiological subtypes through clustering or subtyping analyses on multi-state functional connectivity, few existing approaches are applicable here to accommodate the network topology and unique biological architecture of functional connectivity. To address this issue, we propose an innovative Bayesian nonparametric network-variate clustering analysis to uncover subgroups with homogeneous brain functional network patterns integrating different cognitive states. In light of the existing neuroscience literature, we assume there are unknown state-specific modular structures within functional connectivity and simultaneously impose selection to identify informative network features for defining subtypes within unsupervised learning. To further facilitate practical use, we develop a computationally efficient variational inference algorithm to perform posterior inference with satisfactory estimation accuracy. Extensive simulations show the superior clustering accuracy and plausible result of our method. Applying the method to the landmark Adolescent Brain Cognitive Development (ABCD) study, we successfully establish neurodevelopmental subtypes linked with impulsivity related behavior trait, and identify brain sub-network phenotypes under each state to signal neurobiological heterogeneity.
\end{abstract}

\noindent%

\newpage
\spacingset{1.45}

\section{Introduction} \label{Introduction}

A fundamental challenge in neuroscience is to understand how brain activity is synchronized across large-scale networks. Recent advances in functional magnetic resonance imaging (fMRI) technologies facilitate the characterization of functional connections across the brain, known as functional connectivity (FC). Given emerging evidence suggests that behavior profiles under different populations may arise through detectable network patterns of FC, developing proper analytical frameworks on FC will provide unprecedented opportunities for advancing our understanding of brain functional organization and how it could inform development, aging or psychiatric illnesses.

In this work, we consider functional brain development of the adolescent population. Adolescence is a critical period when cognition and emotion continue to mature, and many psychological disorders may emerge along brain development. Recent empirical studies \citep{foulkes2018studying, goddings2019understanding} have revealed a broad variability in brain functional profiles, indicating that adolescence is a highly fluctuated life stage for neurodevelopment. Though has not been explored yet, dissecting the heterogeneous patterns of brain FC profiles could inform neurodevelopmental subtypes that potentially direct an early detection of mental disorders and assist with timely intervention. In this work, we rely on the data collected from the Adolescent Brain Cognitive Development (ABCD) study (https://abcdstudy.org/). This recent prospective study is the largest one to investigate brain development and adolescent health for children aged 9 to 10 years \citep{casey2018adolescent}. For each participant, fMRI was collected under resting state (RS) to capture intrinsic status, and three cognitive task states including the emotional n-back task (EN-back), the Stop Signal task (SST), and the Monetary Incentive Delay (MID) task to measure working memory, emotion regulation, reward processing, motivation, impulsivity, and impulse control. Under each state, the blood-oxygen-level-dependent (BOLD) signals from fMRI could be aggregated at individual regions of interest (ROIs) over the whole brain under an atlas. By characterizing the temporal correlation between those time series from each pair of ROIs, FC could be constructed for each child under each state as an undirect graph to characterize the brain functional organization under the cognitive status. Integrating FC across states, our goal here is to construct neurodevelopment subtypes based on subject's unique multi-state connectivity signatures and simultaneously dissect brain network profiles that define subtypes.

From an analytical perspective, our problem can be considered as an unsupervised learning based on multi-dimensional network-variate data to uncover latent groups with homogeneity. Though there is a broad literature on clustering methods ranging from heuristic and model-based approaches \citep{sun2019bayesian, li2019tutorial, sinaga2020unsupervised}, few are readily applicable to our case to accommodate the unique network architecture and biological topology of FC. Despite that a few attempts have been made to cluster brain connectivity at a single state to explore data structure or disease subtypes  \citep{lin2018brain, chen2019parsing, sellnow2020biotypes}, heuristic K-means algorithms were adopted in their analyses and they used individual connections as input with  network structure disregarded completely. In some other fields like cancer genomics, clustering has been extended to incorporate feature selection and a potential multi-dimensional fashion \citep{shen2009integrative,kim2006variable,argelaguet2018multi,mo2018fully}. In those modeling frameworks, the observed features are summarized into one or multiple vectors. Though some methods incorporate the relationship among features, such as through latent variables \citep{shen2009integrative} or Markov random field model \citep{zhao2021bayesian}, it is completely different from our setting with network-variate inputs for each subject. From a more graphical learning perspective, \cite{mukherjee2017clustering} proposed a network-valued clustering method based on either graphon or global network properties. However, their method was specifically designed for binary social networks, and cannot be directly implemented to study brain networks with distinct graphic architectures. To the best of our knowledge, only two papers \citep{tokuda2021multiple,dilernia2022penalized} considered clustering brain connectivity without destroying the network structure. However, they both directly modeled the clustering of connectivity matrices via a mixture of Wishart distributions which could suffer from instability due to the large noise within FC entries; and there was also no selection embedded in their model to remove noise features and enhance interpretability.

To fill this gap, we develop a unified Bayesian nonparametric clustering method under multi-state network-variate connectivity features in this work. Instead of directly extracting unique brain connections from each FC, we parameterize connectivity networks via weighted stochastic block models (SBMs) \citep{faskowitz2018weighted} in light of the biological architecture of FC. Specially, converging evidence in neuroscience reveals that brain functional organizations tend to exhibit through a number of sub-networks, or modules \citep{schwarz2008community, ferrarini2009hierarchical}. Under resting state, canonical functional sub-networks have been established to reflect a series of functional systems for the instinct status \citep{shen2010graph}. Some more recent explorations further demonstrate that such modular structures also vary across different cognitive states. In our model, we assume the state-dependent modular structures are unknown, and will be simultaneously uncovered within our learning framework. Using SBMs to characterize brain functional network patterns has started to receive growing attention lately \citep{pavlovic2020multi,zhang2020mixed,ZY2022}. In the current setting, we integrate multi-state SBMs within Bayesian nonparametric clustering, where the modular structures could be informed along the subtyping process with the informative connectivity features defining subtypes identified.

Our contributions in this paper are multi-fold. First, we fill the analytical gap to develop an unsupervised learning framework for network-variate features arising from multi-state brain FC. Building on a Dirichlet process mixture (DPM) model, we define neurobiological subtypes through learning the heterogeneity of brain networks in a nonparametric paradigm. Second, in light of the biological architecture, we incorporate a set of SBMs with unknown community allocations to accommodate the modular structure in the brain functional system. The estimated state-specific modular structure delineates the specific FC pattern driving the observed neurobiological heterogeneity. To further improve interpretability and guide potential brain intervention targets, we
simultaneously select informative network features and exclude noisy ones. Third, to improve practical use, we develop a variational algorithm to accomplish posterior inference, which dramatically reduces the computational cost compared with conventional Markov chain Monte Carlo (MCMC) methods. Finally, we apply the proposed model to the multi-state FC data collected under RS and three cognitive tasks for children in the landmark ABCD study, and obtain meaningful results on neurodevelopmental subtyping.

The rest of the paper is organized as follows. In Section \ref{Methods}, we introduce the network-variate clustering framework, the prior specifications, and the efficient variational inference algorithm of our proposed method, named as \textit{M}ulti-di\textit{M}ensional \textit{B}ay\textit{e}sian nonp\textit{a}rametric \textit{n}etwork \textit{s}ubtyping (MMBeans). In Section \ref{simulation}, we conduct extensive simulations to assess the performance of MMBeans in comparison to existing methods. We further implement MMBeans to the multi-state connectivity data from the ABCD study to investigate neural development subtypes. In Section \ref{Discussion}, we conclude with discussions.

\section{Methods} \label{Methods}

\subsection{Model formulation} 
Suppose $N$ subjects have their brain fMRIs collected over $M$ cognitive states (e.g. resting and performing tasks). Under an MRI atlas, for each state, we partition the brain into a set of nodes, or ROIs as $\mathcal{V}$ with a cardinality $V$.
Across the nodes, FC for subject $i$ at state $m$ can be summarized by an undirected graph $\mathcal{G}_{im}=(\mathcal{V}, \mathcal{E}_{im})$ with $\mathcal{E}_{im}$ denoting the set of brain connections. Over the states, multi-dimensional FC can then be represented by a unique semi-symmetric tensor $\mathcal{A}_i\in \mathbb{R}^{V\times V\times M}$, where the frontal slice $\mathcal{A}_i(:,:,m)$ summarizes the symmetric connectivity matrix at state $m$ uniquely determined by $\mathcal{G}_{im}$. Each element within the connectivity tensor $\mathcal{A}_i(v,v',m)=a_{im,vv'}$ characterizes the connection between ROIs $v$ and $v'$, and it could take different data types including continuous and binary values in practice depending on the way connectivity is summarized.

To construct neurobiological subtypes through clustering on the connectivity-based tensor $\mathcal{A}_i$, it is desirable to regulate our learning by the underlying network geometry while exploiting unique biological architectures for each FC. The latter aspect is particularly crucial to discover subgroups of individuals that share similar neural interconnection patterns in a biologically meaningful way. Specifically, recent neuroscience studies indicate that whole brain functional organizations are composed of a series of modular systems or sub-networks \citep{schwarz2008community,nicolini2016modular}. Under resting state for instance, canonical sub-networks \citep{power2011functional} have been defined to reflect the intrinsic community configurations in human brain function. To accommodate the modular structure, we assume the node set $\mathcal{V}$ can be partitioned into $S_m$ unknown disjoint subsets/communities at state $m$. Instead of prespecifying the modular membership  \citep{zhang2020mixed}, we assume the network communities are unknown and vary across states. Such specifications are based on the following two considerations -- first, existing canonical sub-networks are constructed under resting state via ad hoc community detection, and may not be applicable to integrate with the current learning objective. Second, latest neuroanatomical literature reveals that brain functional suborganizations reconfigure in a meaningful manner dependent upon what the brain is doing \citep{salehi2020there}. This indicates that we will expect different sub-network configurations under different states. Therefore, under each state,  we characterize the partition allocations by indicator matrices $\boldsymbol{Z}_m \in \mathbb{R}^{V\times S_m}$ with its element $\boldsymbol{Z}_m(v,s)=z_{m,vs}$ equal to 1 if node $v$ belongs to community $s$ at state $m$, and 0 otherwise. We then model the latent community indicator vector for node $v$ denoted by $\boldsymbol{z}_{m,v\cdot}$ by an independent and identically distributed (iid) Multinomial distribution: $\boldsymbol{z}_{m,v\cdot} \stackrel{iid}{\sim}\text{Mult}(\boldsymbol{\tau}_m)$,
where $\boldsymbol{\tau}_m=(\tau_{m, s}; 1 \le s \le S_m)$ captures the probabilities of a node allocating into each of the $S_m$ communities, and $\sum_{s} \tau_{m,s}=1$. Conditional on $\{\boldsymbol{Z}_m\}_{m=1}^M$, each 
element of $\mathcal{A}_i$ can be represented generatively via an SBM  
\begin{equation}\label{eq:sbm}
    a_{im,vv'}\mid z_{m,vs}=1, z_{m,v's'}=1 \sim f(a_{im,vv'}; \boldsymbol{\theta}_{im,ss'}),\quad  1\leq s\leq s'\leq S_m,
\end{equation}
where $f(\cdot)$ describes the distribution function for the connectivity metric between blocks $s$ and $s'$ under the parameter set $\boldsymbol{\theta}_{im,ss'}$. Depending on the realization of  $f(\cdot)$, $\boldsymbol{\theta}_{im,ss'}$ could consist of mean and variance parameters in a Normal distribution, or probability parameter in a Bernoulli distribution. Under the assumption of an SBM, for subject $i$ and state $m$, $\boldsymbol{\theta}_{im,ss'}$ will be independent across different blocks for $1 \le s\leq s' \leq S_m$. In other words,  through model (\ref{eq:sbm}),
we manage to segregate the topological information characterized by $\boldsymbol{Z}_m$  from each connectivity matrix $\mathcal{A}_i(:,:,m)$, allowing us to operate on network parameters  $\boldsymbol{\theta}_{im}=(\boldsymbol{\theta}_{im,ss'}; 1\leq s\leq s'\leq S_m)$ to facilitate neurobiological subtyping without a topological constrain.

On the other hand, it is well known that brain functional signals are inherently sparse with the majority of brain alternations operated by a few functional organizations \citep{finn2015functional,drysdale2017resting}. This suggests that a subset of connectivity metrics primarily contribute to defining subtypes, and they will also play a crucial role as the potential intervention targets.  To identify those informative network features and exclude the noise ones along subtyping, we introduce a latent indicator set $\bfgamma=(\gamma_{m,ss'};~ m=1,\dots,M, 1\leq s\leq s'\leq S_m )$, where each of its element $\gamma_{m,ss'}=1$ if connectivity between blocks $s$ and $s'$ distinguishes different subtypes, and $\gamma_{m,ss'}=0$ otherwise. Based on $\gamma_{m,ss'}$, we could represent $f(a_{im,vv'}; \boldsymbol{\theta}_{im,ss'})$ as a mixture distribution
\begin{equation}
\gamma_{m,ss'}f(a_{im,vv'}; \boldsymbol{\theta}^1_{im,ss'})+(1-\gamma_{m,ss'})f(a_{im,vv'}; \boldsymbol{\theta}^0_{m,ss'}),
\end{equation}
where $\boldsymbol{\theta}^1_{im,ss'}$ and $\boldsymbol{\theta}^0_{m,ss'}$ denote the parameter sets for the informative and noise components, respectively. We set $\boldsymbol{\theta}^1_{im,ss'}$ to  vary across subjects to define subtypes and $\boldsymbol{\theta}^0_{m,ss'}$ to be subject-invariant; and further denote $\boldsymbol{\Theta}_i^1=(\boldsymbol{\theta}^1_{im,ss'}; m=1,\dots,M, 1\leq s\leq s'\leq S_m)$ and $\boldsymbol{\Theta}^0=(\boldsymbol{\theta}^0_{m,ss'}; m=1,\dots,M, 1\leq s\leq s'\leq S_m)$. 
Given $\bfgamma$, we assign noninformative priors for $\boldsymbol{\Theta}^0$, for instance, set mean to zero and variance to a large value for a Normal distribution or probability to 0.5 for a Bernoulli distribution. In terms of the informative components,  we consider the following nonparametric Dirichlet process ($\mathcal{DP}$) model to induce clustering
\begin{equation}\label{eq:dp}
    \begin{aligned}
    \boldsymbol{\Theta}^1_{i}\mid \bfgamma, G &\sim G \qquad
G \sim \mathcal{DP}(G_0, \alpha),
    \end{aligned}
\end{equation}
with $G$ a joint probability measure for the parameters in $\boldsymbol{\Theta}^1$, and $G_0$ and $\alpha$ the base measure and scale parameter characterizing the location and degree of concentration for its obtained samples. Model (\ref{eq:dp}) can be further specified using a stick-breaking representation \citep{sethuraman1994constructive} via an infinite weighted sum as
\begin{equation}\label{eq:dp2}
    \begin{aligned}
    G &= \sum_{d=1}^{\infty}w_d\delta_{\Lambda_d}; 
    \quad 
    \text{with }~\Lambda_d\stackrel{iid}{\sim} G_0, \quad 
    w_d  = w'_d\prod_{l<d}(1-w'_l), 
    \quad  w'_l \stackrel{iid}{\sim} \mbox{Beta}(1, \alpha),
    \end{aligned}
\end{equation}
where $\delta_{\Lambda_d}$ is a degenerate probability function on $\Lambda_d$, with each $\Lambda_d$ drawn independently from the base measure $G_0$. Such a representation clearly reveals that distribution $G$ is almost surely discrete with its realization sampling from infinite point masses $\{\Lambda_d\}_{d=1}^{\infty}$ under weights $\{w_d\}_{d=1}^{\infty}$. Accordingly, the sampling for $\boldsymbol{\Theta}^1_{i}$ across the subjects will take values directly from $\{\Lambda_d\}_{d=1}^{\infty}$, and eventually concentrate on a few of its initial components given that the sampling weight $w_d$ decreases exponentially with $d$ increased. Without a need to specify the number of clusters, the proposed modeling framework induces subtyping of subjects in light of the identified informative multi-state network features with each subtype sharing the same values for $\boldsymbol{\Theta}^1$ to describe the subtype-specific connectivity profiles.

\begin{figure}[t]
\centering
\includegraphics[scale=0.45]{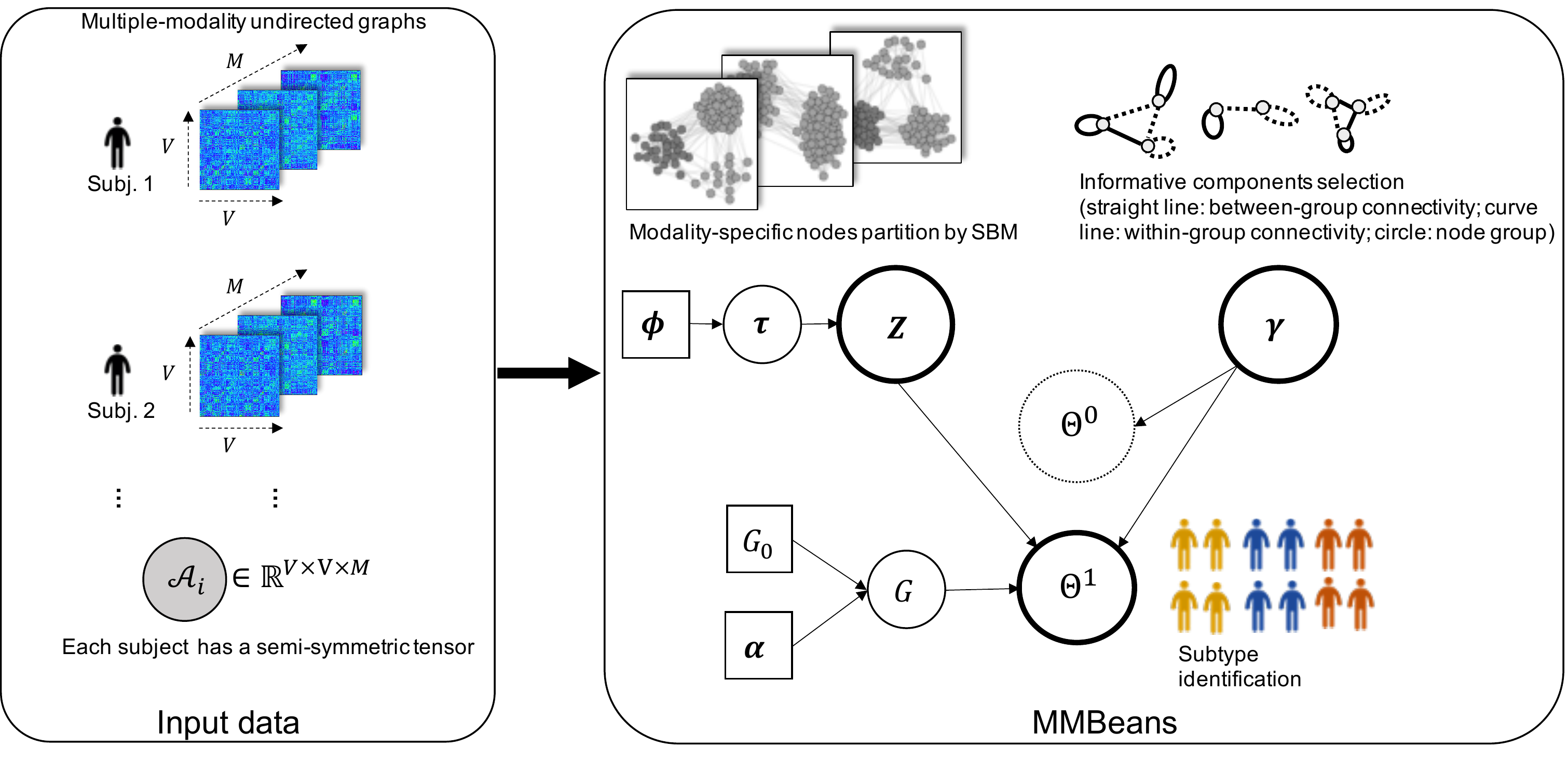}
\caption{\textit{The schematic diagram of MMBeans. With the input of multi-state undirected graphs, MMBeans can simultaneously infer the state-specific modular structure with SBM technique, select informative features, and cluster subjects into subtypes aided further by DPM. Circles represent variables, shaded circle is observed, and squares represent hyperparameters. Non-informative priors are not included in this diagram.}}
\label{diagram}
\end{figure}

In practice, we could determine the realization for $f(\cdot)$ during model implementation based on the input connectivity data. When adopting a Normal distribution function for $f(\cdot)$, we assume base measure $G_0=\prod_m^M\prod_{s=1}^{S_m}\prod_{s'\ge s}\mbox{NIG}(0, \lambda, \frac{\alpha_1}{2}, \frac{\beta_1}{2})$ as a joint distribution combined by Normal-Inverse-Gamma (NIG) distributions corresponding to the mean and variance parameters for each element in $\boldsymbol{\Theta}^1$. 
Similarly, when $f(\cdot)$ is a Bernoulli distribution function, we set $G_0=\prod_m^M\prod_{s=1}^{S_m}\prod_{s'\ge s}\mbox{Beta}(\alpha_0, \beta_0)$ for each of the probability parameter in this informative parameter set.  We also set a noninformative Gamma prior for the $\mathcal{DP}$ scale parameter as $\alpha\sim \mbox{G}(1,1)$ to induce enough flexibility on the concentration of clustering. For the remaining parameters, we assume a noninformative Bernoulli prior for each selection indicator $\gamma_{m,ss'}$, and a Dirichlet prior for the node allocation probabilities $\boldsymbol{\tau}_m \sim \text{Dir}(\boldsymbol{\phi}_m)$, with prespecified $\boldsymbol{\phi}_m=(\phi_{m,s};s=1,\dots,S_m)$. We name our modeling framework MMBeans (Multi-diMensional Bay\textit{e}sian nonp\textit{a}rametric \textit{n}etwork \textit{s}ubtyping), and a demonstration of our modeling scheme, input data structure and outputs is presented in Figure \ref{diagram}.

\subsection{Variational Inference}

To estimate model parameters for MMbeans, we develop a  posterior inference algorithm. To first provide a more feasible computational allocation, following a commonly used strategy \citep{ishwaran2001gibbs,zhao2022bayesian}, we truncate the stick-breaking representation by setting a conservative upper bound $D$ for the possible number of subtypes. By introducing a subtyping membership matrix $\boldsymbol{C}=(\boldsymbol{c}_1,\dots,\boldsymbol{c}_N)^T$ with each latent vector $\boldsymbol{c}_i\in (1, \dots, D)$ capturing the subtype allocation for subject $i$, we have each $\boldsymbol{c}_i$ follow a Multinomial distribution with probabilities $\boldsymbol{w}=(w_1,\dots,w_D)$, and each $w_d$ is determined by $(w'_1,\dots,w'_d)$ as shown in (\ref{eq:dp2}). 
Denote all the unknown parameters in the model as $\boldsymbol{\Xi} = \{ \{\boldsymbol{\tau}_m, \boldsymbol{Z}_m \}_{m=1}^M, \boldsymbol{w}', \boldsymbol{C}, \boldsymbol{\gamma}, \boldsymbol{\Theta}^1, \boldsymbol{\Theta}^0\}$. Based on the posterior likelihood of $\boldsymbol{\Xi}$, one option is to estimate the posterior distribution for each parameter via a Markov Chain Monte Carlo (MCMC) algorithm. With a full conditional distribution derived for each parameter, an MCMC can be performed via Gibbs samplers. However, in practice with high-dimensional feature space, an MCMC could suffer with poor mixing and requires intensive computation for the algorithm to converge. Therefore, we employ an alternative variational inference (VI)  \citep{blei2006variational, ormerod2010explaining} to conduct the posterior estimation for our proposed model.

The main idea of VI is to cast the inference as an optimization problem through seeking a surrogate posterior distribution, known as variational distribution (denoted by $q(\boldsymbol{\Xi})$), that minimizes the similarity with the true posterior. Here, we consider the simple but powerful mean-field approximation by assuming the latent variables are mutually independent. Suppose $\boldsymbol{\Xi}$ can be partitioned into non-overlapping groups and each is denoted by $\Xi_l$, the variational distribution fully factorizes as $q(\boldsymbol{\Xi}) = \prod_{l}q(\Xi_l)$. Our optimization problem is to minimize its Kullback-Leibler (KL) divergence from the true posterior distribution, $p(\boldsymbol{\Xi}|\mathcal{A})$. With simple algebraic manipulation, we can represent the log marginal distribution of the observations as:
\begin{equation} \label{eq:ELBO}
    \log p(\mathcal{A}) = \int q(\boldsymbol{\Xi})\log \{\frac{p(\boldsymbol{\Xi},\mathcal{A})}{q(\boldsymbol{\Xi})}\} d\boldsymbol{\Xi}  + \int q(\boldsymbol{\Xi}) \log \{\frac{q(\boldsymbol{\Xi})}{p(\boldsymbol{\Xi}|\mathcal{A})}\} d\boldsymbol{\Xi},
\end{equation}
where the second term on the right hand side is the KL divergence between $q(\boldsymbol{\Xi})$ and $p(\boldsymbol{\Xi}|\mathcal{A})$, and because of the non-negative KL divergence, the first term denoted as $\mathcal{L}(\boldsymbol{\Xi})$ represents a lower bound for $\log p(\mathcal{A})$.  Therefore, minimizing the KL divergence is equivalent to maximizing the evidence lower bound (ELBO), $\mathcal{L}(\boldsymbol{\Xi})$, which is usually more tractable in practice. Suppose the conditional distribution of each $\Xi_l$ belongs to an exponential family. We have $p(\Xi_l|\boldsymbol{\Xi}_{-l},\mathcal{A}) = h(\Xi_l)\exp\{\eta(\boldsymbol{\Xi}_{-l}, \mathcal{A})T(\Xi_l) - A(\boldsymbol{\Xi}_{-l}, \mathcal{A})\}$, where $\eta(\boldsymbol{\Xi}_{-l}, \mathcal{A})$ is the natural parameter.  For each variational factor, the ELBO maximization occurs when 
\begin{equation}\label{eq:q_update}
\begin{aligned}
q(\Xi_l) = \operatorname*{argmax}_q \mathcal{L} (\Xi_l)
&\propto 
\exp(\mathbb{E}_{-q_l}[\log p(\Xi_l, \boldsymbol{\Xi}_{-l},\mathcal{A})]),
\\
&\propto 
h(\Xi_l)\exp \{ \mathbb{E}_{-q_l}[\eta(\boldsymbol{\Xi}_{-l}, \mathcal{A})] T(\Xi_l),
\end{aligned}
\end{equation}
where $\boldsymbol{\Xi}_{-l}$ denotes all the other latent variables except the $l$-th, and $\mathbb{E}_{-q_l}$ means taking expectation with respect to the variational densities of $\boldsymbol{\Xi}_{-l}$.

The second proportion of (\ref{eq:q_update}) suggests that the optimal variational distribution for each latent variable is actually the same family as the true conditional distribution \citep{blei2017variational}. Therefore, the variational distribution for our model fully factorizes as
\begin{equation} \label{eq:q_factor}
q(\boldsymbol{\Xi}) = \bigl\{ \prod_{m=1}^M q(\boldsymbol{\tau}_m)q(\boldsymbol{Z}_m)\bigr\} q(\boldsymbol{w}'_d)q(\boldsymbol{C})q(\boldsymbol{\gamma})q(\boldsymbol{\Theta}^1)q(\boldsymbol{\Theta}^0),
\end{equation}
where we assume each factor $q(\cdot)$ belongs to an exponential family. As stated previously, when each brain connection is summarized continuously,  we have $\boldsymbol{\Theta}^1:=(\mu_{dm,ss'},\sigma^2_{dm,ss'}; d=1,\dots,D, m=1,\dots,M, 1\leq s\leq s'\leq S_m)$ consisting of mean and variance parameters in the Normal distribution for each informative component. When the connectivity weight is dichotomized, we have 
$\boldsymbol{\Theta}^1:=(\rho_{dm,ss'}; d=1,\dots,D, m=1,\dots,M, 1\leq s\leq s' \leq S_m)$ collecting the probability parameters.
Under $q(\cdot)$, we propose the following  realizations on each of the variational distributions
\begin{equation}\label{eq:q_dist}
    \begin{aligned}
        \boldsymbol{\tau}_m 
        & \sim \text{Dir}(\boldsymbol{t}_m),\quad \boldsymbol{t}_m=(t_1, \dots, t_{S_m}), \quad m=1,\dots,M;
        \\
        \boldsymbol{z}_{m,v\cdot} 
        & \sim \text{Mult}(\boldsymbol{\eta}_{m,v}),\quad 
        \boldsymbol{\eta}_{m,v}=(\eta_{m,v1},\dots, \eta_{m,vS_m}),\quad m=1,\dots,M, \quad v=1,\dots,V;
        \\
        w'_d 
        & \sim \text{Beta}(e_d, f_d), \quad
        d=1,\dots,D, \quad
        \text{with } ~q(w'_D=1)=1;
        \\
        \boldsymbol{c}_i 
        & \sim \text{Mult}(\frac{e^{b_{i1}}}{\sum_{l}e^{b_{il}}}, \dots, \frac{e^{b_{iD}}}{\sum_{l}e^{b_{il}}}), \quad
        i=1,\dots,N;
        \\
        \gamma_{m,ss'} 
        & \sim \text{Bern}(\text{expit}(\zeta_{m,ss'})),\quad
        m=1,\dots,M, \quad 1 \leq s \leq s' \leq S_m;
        \\
         \sigma^2_{dm,ss'} 
        & \sim \text{IG}(\frac{g_{dm,ss'}}{2}, \frac{h_{dm,ss'}}{2}), ~
        \mu_{dm,ss'}\mid\sigma^2_{dm,ss'} 
        \sim \text{N}(u_{dm,ss'},\frac{\sigma^2_{dm,ss'}}{r_{dm,ss'}}) \text{ or }~  
        \\
        \rho_{dm,ss'} &\sim \mbox{Beta}(j_{dm,ss'}, k_{dm, ss'}),
        \quad d=1,\dots, D, \quad m=1,\dots,M, \quad 1\leq s \leq s' \leq S_m.
    \end{aligned}
\end{equation}
where $\text{expit}(x)=1/(1+\exp(-x))$ represents the logistic sigmoid function, and depending on the realization of $f(\cdot)$, we will impose corresponding variational distributions for Normal or Bernoulli parameters.

The cyclic dependencies shown in (\ref{eq:q_update}) suggest an iterative coordinate ascent algorithm, and we will be able to obtain closed-form updates under our distribution choices by (\ref{eq:q_dist}). The detailed update equations in the variational algorithm are provided in the Supplementary material available online, and we briefly summarize each step in Algorithm \ref{alg}. Under random initializations, the algorithm repeatedly updates each variational parameter until the change in ELBO values shows a convergence. Then the subject clusters, sub-network structure and informative blocks selection can be inferred by the variational parameters $\{b_{id}, i=1,\dots,N, d=1,\dots,D\}$, $\{\boldsymbol{\eta}_{m,v}, m=1,\dots,M, v=1,\dots,V\}$ and $\{\zeta_{m, ss'}, m=1,\dots,M, 1 \leq s \leq s' \leq S_m \}$ respectively. In particular, suppose the largest weight for $q(\boldsymbol{c}_i)$ comes from $b_{id}$, subject $i$ then belongs to cluster $d$. Similarly, the node $v$ from state $m$ is assigned to block $s$ if $\eta_{m, vs}$ is the largest among $\boldsymbol{\eta}_{m, v}$. The connectivity between sub-networks $s$ and $s'$ from state $m$ is discriminative when $\text{expit}(\zeta_{m,ss'})$ is greater than 0.5. Otherwise, this connectivity is considered as noise.

\begin{algorithm}[!htb]
  \caption{Variational Inference Algorithm for MMBeans}\label{alg}
  \begin{algorithmic}
  \State Input Data: multi-state FC tensor $\mathcal{A}$.
  \State Initialize the variational parameters for the variational distributions of \\  
  $\boldsymbol{\Xi} =
  \{ \{\boldsymbol{\tau}_m, \boldsymbol{Z}_m \}_{m=1}^M, \boldsymbol{w}', \boldsymbol{C}, \boldsymbol{\gamma}, \boldsymbol{\Theta}^1\}$.
      \While{Convergence is not reached}:
        \State Sequentially update the following variational parameters with equations provided in \\ 
        the Supplementary material available  online. 
        \State For $1 \le d < D$, update $e_d$, $f_d$ by (S3).
        \State For $1\le m \le M$, $1\le d \le D$ , and $1\leq s\leq s'\leq S_m$, update $g_{dm,ss'}$, $r_{dm,ss'}$, $u_{dm,ss'}$, $h_{dm,ss'}$  by (S4), or update $j_{dm,ss'}$, $k_{dm, ss'}$ by (S10) according to input data type.
        \State For $1\le m \le M$, and $1\le s\le s'\le S_m$,  update $\zeta_{m,ss'}$ by (S5) or (S11) according to input data type.
        \State For $1\le i \le N$ and $1\le d \le D$, update $b_{id}$ for $1\le d \le D$ by (S6) or (S12) according to input data type.
        \State For $1\le m\le M$, and $1\leq s\leq S_m$, update $t_{m,s}$ by (S7).
        \State For $1\le m \le M$, $1\le v\le V$, and $1\le s \le S_m$, update $\eta_{m,vs}$ by (S8) or (S13) according to input data type. 
      \EndWhile\label{euclidendwhile}
  \end{algorithmic}
\end{algorithm}

\section{Simulation study} \label{simulation}
\subsection{Simulation design} 
\begin{figure}[!t]
\centering
\includegraphics[width=0.9\linewidth]{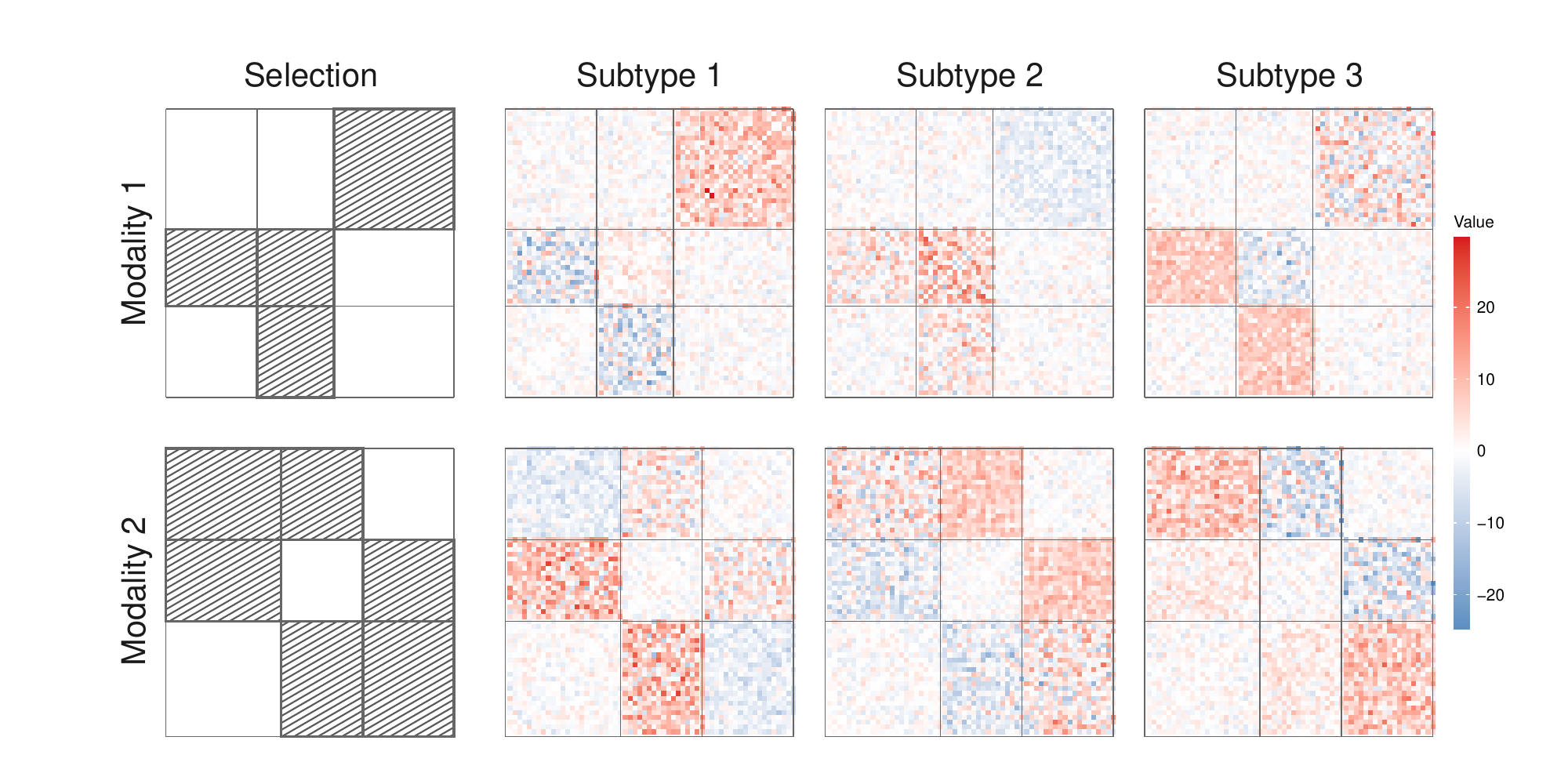}
\caption{\textit{Example of simulation data with two states and three subject clusters. The modular structure is displayed with block pattern. Each row represents one state. The first column shows the feature selection pattern with the shaded block being regarded as informative ones. The right three columns show the patterns of simulation data for three clusters with red and blue colors indicating positive and negative values.}}
\label{sim_data_block}
\end{figure}

We evaluate the finite sample performance of the proposed MMbeans model using simulations. We set $N=100$ with two cognitive states collected for each subject. Among all the subjects, we assume there are three neurobiological subtypes defined by the multi-state FC, and we randomly allocate subjects into the three subtypes with equal probabilities. To generate FC, we vary the number of nodes $V=60, 200$ and $500$ to cover possible sizes of the currently used brain atlases \citep{glasser2016multi,shen2013groupwise}. Under each connectivity dimension, we randomly partition the nodes into three state-dependent modules by Multinomial distributions with the corresponding probabilities equal to $(0.25, 0.40, 0.35)$ and $(0.30, 0.30, 0.40)$ for each state, respectively. These lead to six unique modular components under each state, and we assume half of them are the informative ones to define subtypes with locations generated randomly under each state. We work on  continuous scales for connectivity matrices in this case. To specify the modular parameters in $f(\cdot)$,  for the informative ones, we generate the Normal distribution means from ${(-3, 2, 7)}$ and variances from $(3, 5, 7)$ under each subtype; and for the noisy elements which are subtype-independent, we set their means to zero and variances to be either 6 or 10 corresponding to a high signal-to-noise ratio (SNR) and a low SNR setting. We show in Figure \ref{sim_data_block} one of the simulated settings with $V=60$ to exemplify modular structures and network effects among different neurobiological subtypes. We generate $50$ Monte Carlo datasets for each simulated setting. 

To implement MMBeans, we set $\lambda=1, \alpha_1=10, \beta_1=10$ for the hyperparameters relating to $G_0$, and employ a flat Dirichlet prior ($\phi_{m,s}=1$) for nodes allocation, however, 
the size of modular $(S_m; m=1, \dots, M)$ for each modality needs to be tuned. We select the optimal $S_m$ based on variational Bayesian information criterion (VBIC) proposed by \cite{you2014variational}, which is defined as $-2\mathbb{E}_q(\log p(\mathcal{A}|\boldsymbol{\Xi})) + 2\mathbb{E}_q(\log q(\boldsymbol{\Xi}))$ and functions similarly as BIC. Given few existing unsupervised learning methods can handle network- or matrix-variate input, to apply them to our multi-state connectivity data, we have to extract the unique connections as their inputs by vectorizing the connectivity matrices. We compare our method with a few competing clustering methods including a very recent Bayesian clustering method Nebula \citep{zhao2021bayesian}, a widely used frequentist latent variable clustering method iCluster \citep{shen2009integrative}, and the canonical heuristic  K-means algorithm (KML). 
Similar to our proposed MMbeans, both Nebula and iCluster are capable of performing  clustering integrating multi-state data with feature selection embedded. To identify informative features along KML, after clustering the subjects, we fit a Multinomial logistic regression model for cluster labels under all the connections and impose a lasso penalty to perform feature selection. The model implementation for Nebula and iCluster follow closely to their recommended setups with the number of subtypes determined along the posterior inference for Nebula and by minimizing the proportion of deviance for iCluster, respectively. As for KML, the cluster number is determined by maximizing Silhouette distance \citep{rousseeuw1987silhouettes}, while the lasso regulating parameter is chosen via 5-fold cross validation.

To evaluate the performance of different methods, we focus on the following aspects. First, we assess subtyping accuracy via the Adjusted Rand Index (ARI) \citep{hubert1985comparing}, which is equal to 1 only if the estimated clustering is identical to the actual one and close to 0 for random partition. Second, we evaluate the performance of distinguishing informative and non-informative network features using sensitivity (sen), specificity (spe) and Youden's index (Y-index) proposed by \cite{youden1950index}. As a composite measure, Y-index is defined as $\mbox{sen} + \mbox{spe} - 1$ to characterize the overall selection accuracy. Of note, to ensure consistency among different methods, we always map the selected network components to the original $V\times V\times M$ network tensor scale when calculating each selection metric. In addition, given the proposed MMBeans model is capable to dissect the modular structure within connectivity, we will also check the accuracy of our model to uncover sub-network structures under each modality using ARI by comparing our estimated state-specific parcellation with the ground truth. Finally, we report the computational time for each method to demonstrate their practical feasibility. 

\begin{table}[t]
\centering
\caption{\textit{Subject clustering and feature selection simulation results for MMBeans, Nebula, KML and iCluster are summarized by ARI, sensitivity, specificity and Y-index averaged over $50$ simulations (standard deviation in the parentheses). The best performance for subject clustering under each setting is bolded. The averaged running time of one simulation for each method is recorded in seconds.}}
\label{tab:sim_summary_landscape}
\resizebox{\textwidth}{!}{%
\begin{tabular}{rrrlrrrrlrrr}
\hline
 &  & \multicolumn{1}{c}{\textbf{Subtyping}} &  & \multicolumn{4}{c}{\textbf{Feature selection}} &  & \multicolumn{2}{c}{\textbf{Modular structure}} &  \\ \cline{3-3} \cline{5-8} \cline{10-11}
\multicolumn{1}{c}{\textbf{SNR}} & \multicolumn{1}{c}{\textbf{Method}} & \multicolumn{1}{c}{\textbf{ARI}} &  & \multicolumn{1}{c}{\textbf{Sen}} & \multicolumn{1}{c}{\textbf{Spe}} & \multicolumn{1}{c}{\textbf{Y-index}} & \multicolumn{1}{c}{\textbf{AUC}} &  & \multicolumn{1}{c}{\textbf{\begin{tabular}[c]{@{}c@{}}M1: \\ ARI\end{tabular}}} & \multicolumn{1}{c}{\textbf{\begin{tabular}[c]{@{}c@{}}M2: \\ ARI\end{tabular}}} & \multicolumn{1}{c}{\textbf{\begin{tabular}[c]{@{}c@{}}Running \\ time\end{tabular}}} \\ \hline
 &  &  &  &  &  & $V=60$ &  &  &  &  &  \\
\multirow{4}{*}{High} & MMBeans & \textbf{0.95 (0.15)} &  & 0.83 (0.02) & \textbf{1.00 (0.00)} & \textbf{0.83 (0.02)} & 1.00 (0.00) &  & 1.00 (0.00) & 1.00 (0.00) & 1.31 \\
 & Nebula & 0.91 (0.15) &  & \textbf{0.99 (0.00)} & 0.57 (0.04) & 0.56 (0.04) & — &  & — & — & 14.23 \\
 & KML & 0.89 (0.19) &  & 0.07 (0.04) & \textbf{1.00 (0.00)} & 0.07 (0.04) & — &  & — & — & 1.99 \\
 & iCluster & 0.55 (0.00) &  & 0.87 (0.12) & 0.49 (0.43) & 0.35 (0.32) & — &  & — & — & 993.57 \\
 &  &  &  &  &  &  &  &  &  &  &  \\
\multirow{4}{*}{Low} & MMBeans & \textbf{0.95 (0.15)} &  & 0.83 (0.02) & \textbf{1.00 (0.00)} & \textbf{0.83 (0.02)} & 1.00 (0.00) &  & 1.00 (0.00) & 1.00 (0.00) & 1.21 \\
 & Nebula & 0.85 (0.21) &  & \textbf{0.99 (0.00)} & 0.00 (0.00) & -0.01 (0.00) & — &  & — & — & 12.89 \\
 & KML & 0.86 (0.20) &  & 0.06 (0.04) & \textbf{1.00 (0.00)} & 0.06 (0.04) & — &  & — & — & 1.96 \\
 & iCluster & 0.55 (0.00) &  & 0.83 (0.11) & 0.49 (0.34) & 0.32 (0.23) & — &  & — & — & 981 \\ \hline
 &  &  &  &  &  & $V=200$ &  &  &  &  &  \\
\multirow{4}{*}{High} & MMBeans & \textbf{0.92 (0.17)} &  & 0.83 (0.02) & \textbf{1.00 (0.00)} & \textbf{0.83 (0.02)} & 1.00 (0.00) &  & 1.00 (0.00) & 1.00 (0.00) & 12.74 \\
 & Nebula & 0.89 (0.16) &  & \textbf{0.99 (0.00)} & 0.52 (0.03) & 0.51 (0.03) & — &  & — & — & 378.33 \\
 & KML & 0.82 (0.22) &  & 0.01 (0.00) & \textbf{1.00 (0.00)} & 0.01 (0.00) & — &  & — & — & 22.78 \\
 & iCluster & — &  & — & — & — & — &  & — & — & — \\
 &  &  &  &  &  &  &  &  &  &  &  \\
\multirow{4}{*}{Low} & MMBeans & \textbf{0.92 (0.17)} &  & 0.84 (0.00) & \textbf{1.00 (0.00)} & \textbf{0.84 (0.00)} & 0.97 (0.01) &  & 1.00 (0.00) & 1.00 (0.00) & 9.49 \\
 & Nebula & 0.74 (0.25) &  & \textbf{1.00 (0.00)} & 0.00 (0.00) & -0.00 (0.00) & — &  & — & — & 455.33 \\
 & KML & 0.82 (0.22) &  & 0.01 (0.00) & \textbf{1.00 (0.00)} & 0.01 (0.00) & — &  & — & — & 22.75 \\
 & iCluster & — &  & — & — & — & — &  & — & — & — \\ \hline
 &  &  &  &  &  & $V=500$ &  &  &  &  &  \\
\multirow{4}{*}{High} & MMBeans & \textbf{1.00 (0.00)} &  & 0.84 (0.00) & \textbf{1.00 (0.00)} & \textbf{0.84 (0.00)} & 1.00 (0.00) &  & 1.00 (0.00) & 1.00 (0.00) & 96.43 \\
 & Nebula & 0.86 (0.19) &  & \textbf{0.99 (0.00)} & 0.52 (0.03) & 0.51 (0.03) & — &  & — & — & 5558.97 \\
 & KML & 0.84 (0.21) &  & 0.00 (0.00) & \textbf{1.00 (0.00)} & 0.00 (0.00) & — &  & — & — & 213.76 \\
 & iCluster & — &  & — & — & — & — &  & — & — & — \\
 &  &  &  &  &  &  &  &  &  &  &  \\
\multirow{4}{*}{Low} & MMBeans & \textbf{0.98 (0.09)} &  & 0.81 (0.03) & \textbf{1.00 (0.00)} & \textbf{0.81 (0.03)} & 1.00 (0.00) &  & 1.00 (0.00) & 1.00 (0.00) & 67.35 \\
 & Nebula & 0.76 (0.24) &  & \textbf{1.00 (0.00)} & 0.00 (0.00) & -0.00 (0.00) & — &  & — & — & 5699.18 \\
 & KML & 0.86 (0.20) &  & 0.00 (0.00) & \textbf{1.00 (0.00)} & 0.00 (0.00) & — &  & — & — & 206.1 \\
 & iCluster & — &  & — & — & — & — &  & — & — & — \\ \hline
\end{tabular}%
}
\end{table}

\subsection{Simulation results} 
The simulation results are summarized in Table \ref{tab:sim_summary_landscape} for all the methods under each simulation setting. As shown in the table, our method consistently outperforms the competing methods with respect to both subtyping and selecting connectivity features to define subtypes. Specifically, the proposed MMBeans obtains the highest ARI under all the simulated sample sizes and noise levels, indicating its superiority to separate subjects based on the multi-state connectivity profiles. With the highest Y-index in all the settings compared with competing methods, MMBeans further shows a strong feature selection power for this unsupervised learning framework. In addition, as a unique output, MMBeans simultaneously uncovers the network modular structure under each state. As the corresponding ARIs are exactly one for both states under all simulation settings, it indicates that MMBeans can fully dissect the underlying modular architectures correctly under our current simulation settings. When it comes to different settings, our method maintains a robust performance despite reduced SNRs and network dimensions, promoting its use in real practice with a relevant high noise and wide range of connectivity sizes. In terms of competing methods, Nebula shows a better performance compared with the remaining ones under low and moderate dimensional cases, but deteriorates substantially when feature space and noise level get higher. The KML approach, though achieving a reasonable performance on clustering, fails to identify informative features leading to low Y-index. Finally, it is worth noting the computational complexity to implement each method. As shown by the running time in Table \ref{tab:sim_summary_landscape}, while iCluster is computationally prohibited when $V=\{200, 500\}$, our MMBeans requires a very small computational cost  even in the presence of high dimensionality. This is highly impressive considering the complexity of our Bayesian modeling. The computational intensity for MMBeans is also much lower compared with competing ones including the heuristic KML. 
In summary, the simulation studies demonstrate the power and robustness of our method in defining subtypes and uncovering sub-network structures. The low computational cost further ensures its feasibility in practice.

\section{Real Data Application} \label{application}

\begin{figure}[p]
\centering
\includegraphics[width=0.85\linewidth]{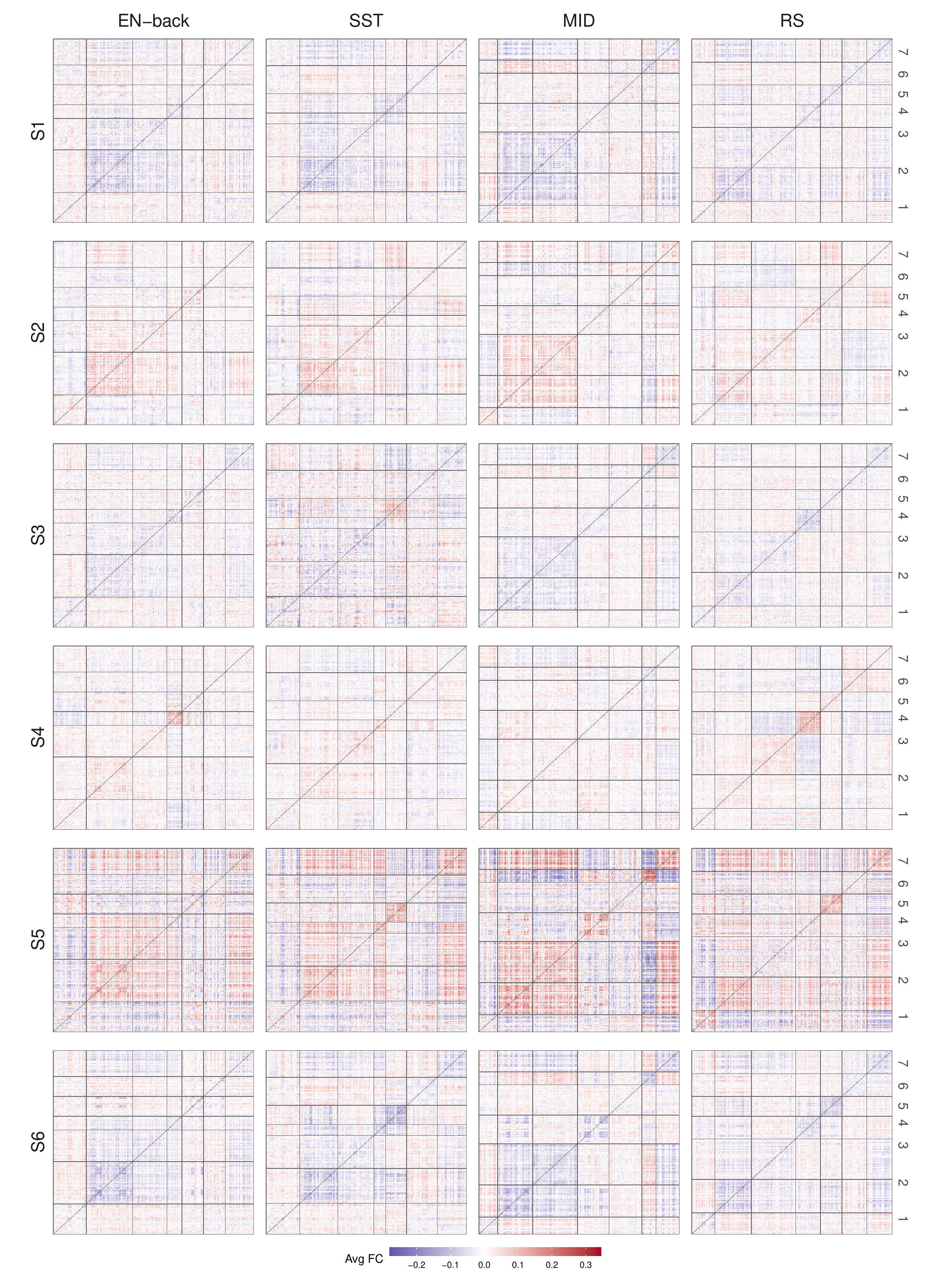}
\caption{\textit{Visualization of scaled averaged FC matrix across subtypes under EN-back, SST, MID and RS. For each combination of imaging condition and subtype, the averaged FC matrix is substracted by condition-specific mean.}}
\label{total_minus_mean_ppH}
\end{figure}

We apply the proposed model to the abovementioned ABCD study. This ongoing landmark children's study plans to collect information from each participant including brain imaging, biospecimen and mental health conditions periodically for ten years. Here, we focus on the first release of fMRI data collected under RS and three task states (EN-back, SST and MID). The details of the imaging acquisition process can be found in \cite{casey2018adolescent}. The raw dicom images for 5,772 subjects were obtained via ABCD fast track (April 2018), and preprocessed using BioImage Suite \citep{joshi2011unified}. The standard preprocessing procedures, such as slice time and motion correction, registration to the MNI template, were described in detail elsewhere \citep{greene2018task, horien2019individual}. The eligible subjects are those scanned under all four states and having qualified scans with no more than 0.10 mm mean frame-to-frame displacement. Finally, 873 subjects were included in our current analysis with the complete data. To construct multi-state FC for each subject, we adopted a 268-node brain atlas \citep{shen2013groupwise} to define ROIs, which includes the cortex, subcortex, and cerebellum. After computing the mean time course across all the voxels within a region, a Pearson’s correlation between the mean time course from each pair of ROIs was computed and scaled to be Normally distributed by a Fisher's Z transformation. After data preparation, our input connectivity tensor for each subject becomes $\mathcal{A}_i\in \mathbb{R}^{268\times 268\times 4}$, and the implementation details for the MMbeans including hyperparameter specifications and the model selection criterion directly follow those in the simulations.

\subsection{Data analysis results}

\begin{figure}[t]
\centering
\includegraphics[width=0.8\linewidth]{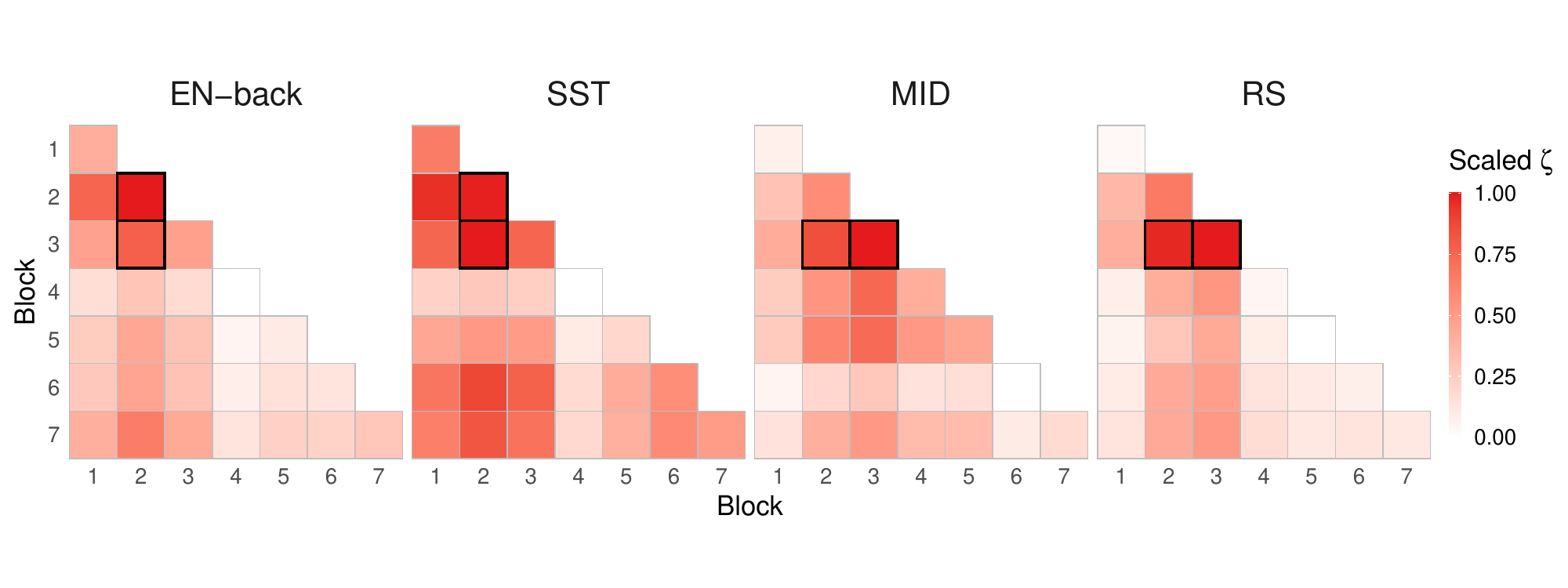}
\caption{\textit{Scaled $\zeta$ value for all pairs of blocks. The two prime features from each state are highlighted with black frames. Darker color shaded cell indicates higher $\zeta$ value.}}
\label{task_N4_cbase}
\end{figure}

\begin{figure}[t]
\centering
\includegraphics[width=0.9\linewidth]{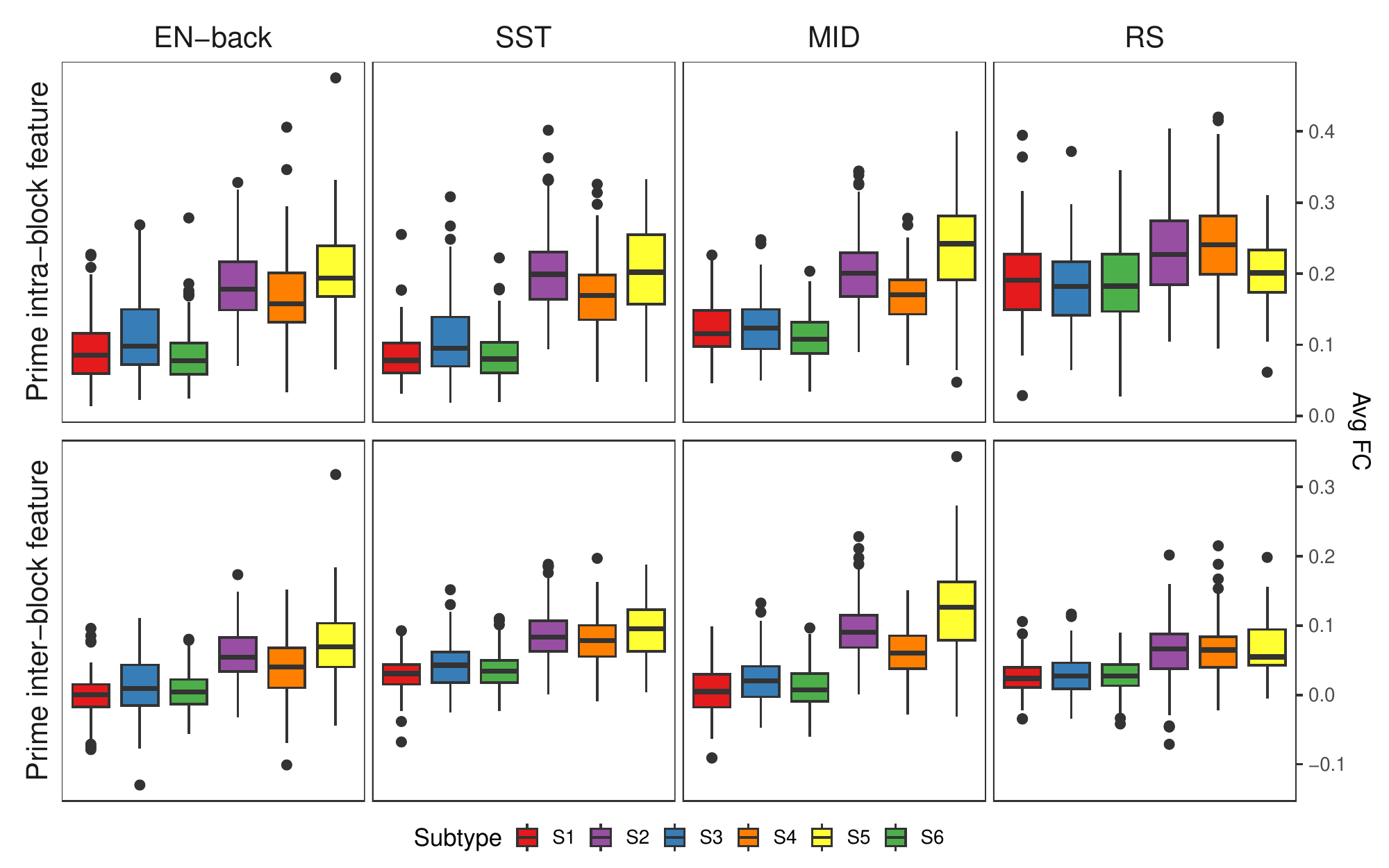}
\caption{\textit{Boxplot of averaged FC of prime features  under EN-back, SST, MID and RS across six subtypes. The averaged FC is calculated by taking the mean of all unique connectivity located in each prime feature for each state and for each subject.}}
\label{total_top2_boxplot}
\end{figure}

\begin{figure}[p]
\captionsetup[subfigure]{justification=centering}
\centering
   \begin{subfigure}[t]{0.9\textwidth}
   \centering
   \includegraphics[width=0.85\linewidth]{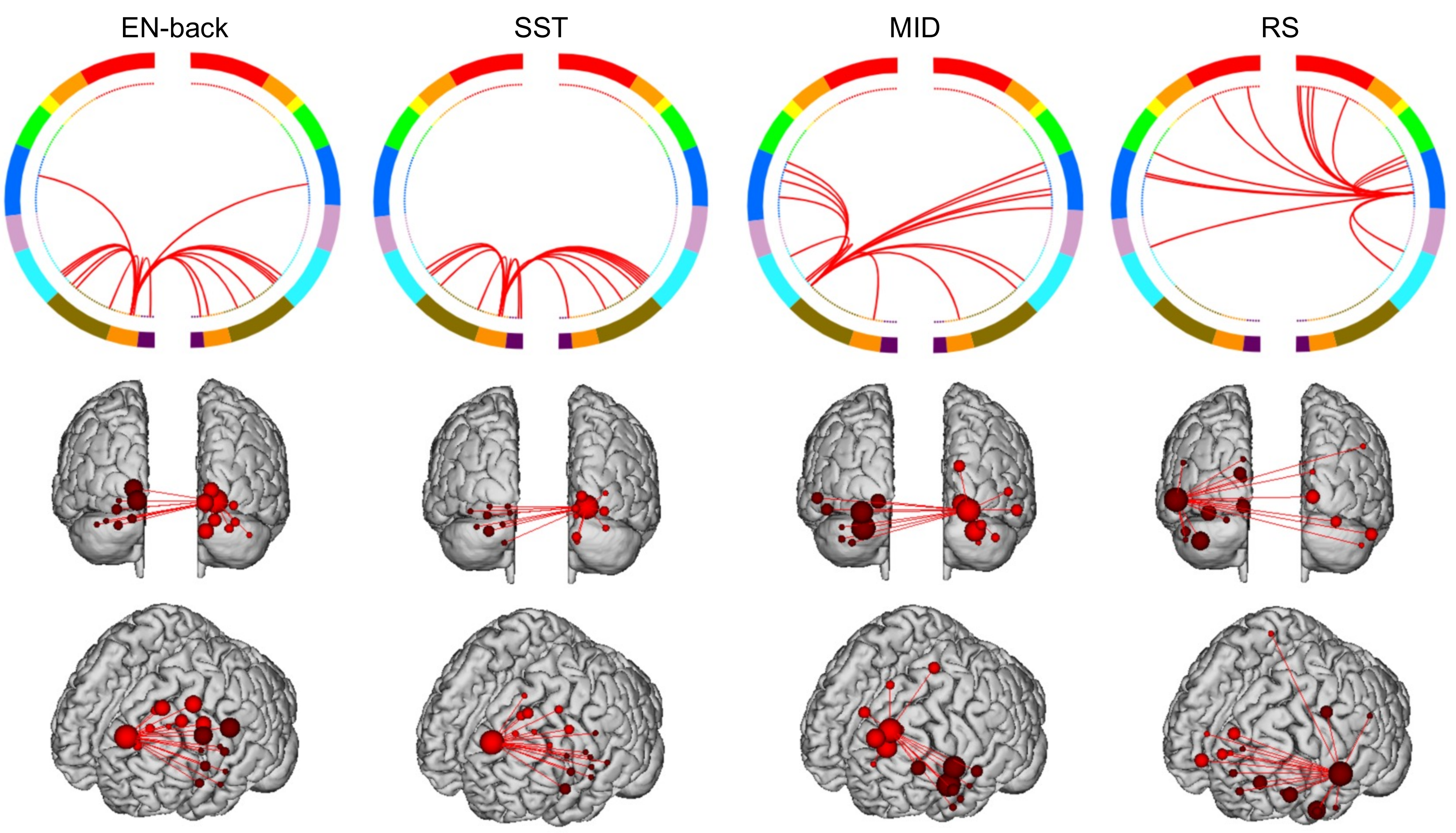}
   \caption{\textit{Top selected connectivity from prime intra-block feature}}
   \label{fig:Ng1} 
   \end{subfigure}

   \begin{subfigure}[t]{0.9\textwidth}
   \centering
   \includegraphics[width=0.85\linewidth]{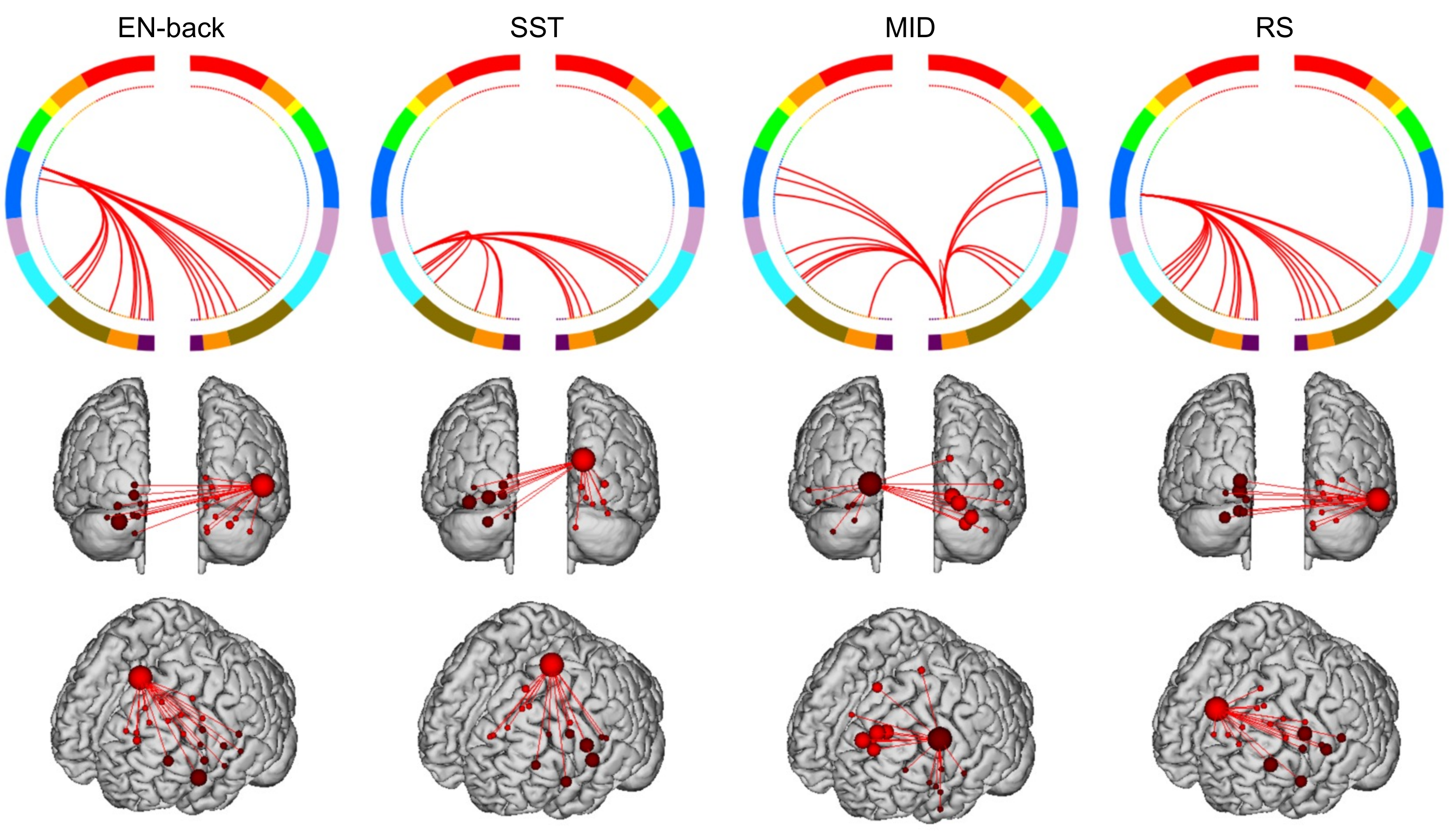}
   \caption{\textit{Top selected connectivity from prime inter-block feature}}
   \label{fig:Ng2}
   \end{subfigure}
   
   \begin{subfigure}[t]{0.9\textwidth}
   \centering
   \includegraphics[width=0.8\linewidth]{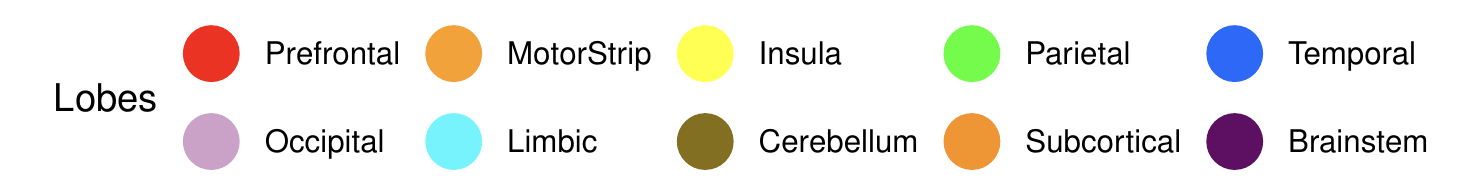}
   \label{fig:legend} 
   \end{subfigure}
\caption{\textit{The location and distribution of the top 5\% most differentiating FC (depicted as red connection lines) from the prime (A) inter- and (B) intra-block features in brain. The circle plots display nodes grouped according to anatomical location, and the 3D brain plots show the back and side views. To perform the selection, we first compare the strength of each connectivity among six subtypes using one-way ANOVA in a post-hoc manner. And only the top 5\% most significant FC from each prime feature is selected to represent the corresponding feature.}}\label{total_top2_connectivity}
\end{figure}

We eventually establish six neurodevelopmental subtypes (indicated as $\mathcal{S}1$ to $\mathcal{S}6$) and partition whole brain ROIs into seven state-specific modular blocks. To understand those heterogeneous brain functional architectures, we first provide the mean-centered averaged FC across subtypes under each cognitive state in Figure \ref{total_minus_mean_ppH}. As shown in the figure, the FC matrix displays a clear block pattern, which supports our modeling for FC in light of modules. In Figure \ref{task_N4_cbase}, we display how strong the connectivity between and within our defined modules contributes to subtyping with a larger $\zeta$ value indicating a higher probability to discriminate subtypes. The two most discriminating features of each state are connectivity within block 2 and between blocks 2\&3 of EN-back and SST, within block 3 and between blocks 2\&3 of MID and RS. For simplicity, we name the connectivity within block 2 or 3 as prime intra-block feature, and connectivity between blocks 2\&3 as prime inter-block feature. Focusing on those most informative features, Figure \ref{total_top2_boxplot} shows how the subtype profiles are defined by multi-state prime features. And Figure \ref{total_top2_connectivity} shows the main brain regions involved in the prime features under each state. Based on the figures, we can see that subtypes $\mathcal{S}1$, $\mathcal{S}3$ and $\mathcal{S}6$ have low average neural activation in two prime features over four states; subtype $\mathcal{S}2$ has high activation in both prime features over four states; $\mathcal{S}4$ have medium activation in task-based prime features and high activation in RS prime features; $\mathcal{S}5$ have high activation in task-based prime features and medium activation in RS prime features.

\begin{figure}[hp]
\centering
\includegraphics[width=0.9\linewidth]{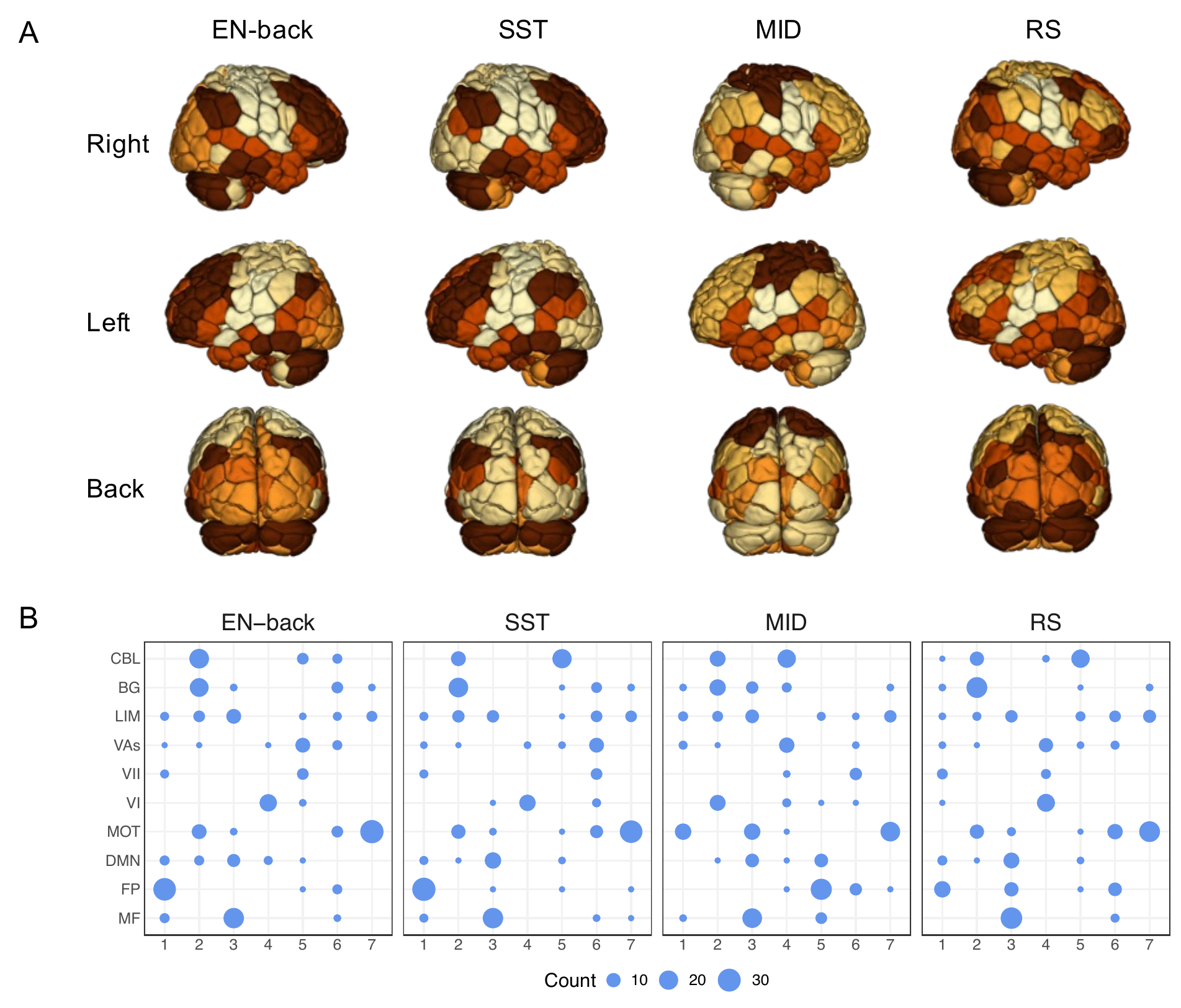}
\caption{\textit{(A) Visualization of algorithm-computed state-specific brain modular structure. 
Brain regions within the same module are colored the same. From top to bottom is the right, left and back view of 3D brain. (B) The association between the yielded block memberships and the canonical resting-state functional sub-networks labels for EN-back, SST, MID, and RS. The point size is scaled to the counts of overlapped nodes between two types of sub-networks. MF: Medial-Frontal, FP: Fronto-parietal, DMN: Default Mode, MOT: Motor, VI: Visual I, VII: Visual II, VAs: Visual Association, LIM: Limbic, BG: Basal Ganglia, CBL: Cerebellum. }}
\label{task_N4_3D_modular_shen10}
\end{figure}

We further examine those dissected modular structures for each state. As shown in Figure \ref{task_N4_3D_modular_shen10}A, our method learned state-specific modular structures accounting for the functional disparities intrigued among different cognitive states. Figure \ref{task_N4_3D_modular_shen10}B further presents a comparison between our subtyping-induced functional modules with the canonical resting-state functional sub-networks \citep{shen2010graph}. As expected, there exists consistency and disparity of node partitions under different states. For instance, block 7 of all four states is mostly located in the MOT sub-network; block 4 of MID, block 1 of EN-back, RS, and SST are mainly within FP sub-network; and both block 4 of EN-back and SST are analogs to VI sub-network. It naturally occurs that the sub-networks with strong primary functions, such as visual sub-network, tend to stay in fewer blocks by our analysis; while those with mixed functions, such as LIM sub-network involved in controlling emotions, motivation, memory and learning \citep{loveland2008fronto,  rajmohan2007limbic, rolls2019cingulate} have been assigned into more blocks. 
In terms of the selected informative blocks, under EN-back, SST and MID, the selected ones are overlapped with MF-CBL-BG, MF-BG-DMN, MF-BG-MOT sub-networks, respectively; and under RS, the selected ones are overlapped with MF-BG-DMN.
Consistent with existing findings, these sub-networks or functional systems have been shown intensively as the primary brain functional features contributing to subtyping under both healthy and disease cohorts \citep{assaf2010abnormal, finn2015functional,chen2016multivariate,drysdale2017resting,rabellino2018cerebellum}.

\begin{table}[t]
\centering
\caption{\textit{The significant clinical and demographic characteristics associated with computed subtypes. Continuous variables are summarized with means and standard deviations, and categorical variables are summarized with counts and proportions.}}
\label{tab:clinical_outcome}
\resizebox{0.9\textwidth}{!}{%
\begin{tabular}{@{}lrrrr@{}}
\toprule
\textbf{Characteristics} & \textbf{All (N=873)} & \textbf{$\mathcal{S}1$ (N=105)} & \textbf{$\mathcal{S}2$ (N=269)} & \textbf{$\mathcal{S}3$ (N=85)} \\ \midrule
UPPS - positive urgency & -0.21 (0.89) & -0.07 (0.93) & -0.30 (0.82) & -0.18 (0.91) \\
Sex: female (vs. male) & 487 (55.85\%) & 57 (54.29\%) & 164 (60.97\%) & 42 (49.41\%) \\
Race: white (vs. other) & 637 (73.39\%) & 68 (65.38\%) & 218 (81.65\%) & 54 (63.53\%) \\ \midrule
 & \textbf{$\mathcal{S}4$ (N=140)} & \textbf{$\mathcal{S}5$ (N=53)} & \textbf{$\mathcal{S}6$ (N=221)} & \textbf{\textit{P}-value} \\ \midrule
UPPS - positive urgency & -0.27 (0.80) & -0.33 (0.90) & -0.09 (0.98) & 0.038 \\
Sex: female (vs. male) & 67 (47.86\%) & 23 (43.40\%) & 134 (60.91\%) & 0.036 \\
Race: white (vs. other) & 120 (85.71\%) & 44 (83.02\%) & 133 (60.73\%) & 0.002 \\ \bottomrule
\end{tabular}}
\end{table}

Finally, we explore how the identified subtypes acquire clinical utility. As suggested by \cite{heeringa2020guide}, we reweight the subjects based on their propensity scores to attenuate the potential selection bias during the sampling and enrollment in our evaluation. We use multilevel linear or logistic models to accommodate variation in data acquisition sites when associating the constructed neurodevelopmental subtypes with behavior and demographic variables. Specifically, we look for behavior traits related to cognitive factors, parental substance use and behavior assessment questionnaires that are of interest to link with neurodevelopment. As shown in Table \ref{tab:clinical_outcome}, we detect significant associations of our subtyping with 
the positive urgency score under Urgency-Premeditation-Perseverance-Sensation Seeking (UPPS) Impulsive Behavior Scale, sex and race; and a complete set of results are provided in the the Supplementary material available online. Based on the results, subtypes $\mathcal{S}1$ and $\mathcal{S}6$ generally have higher positive urgency scores, which characterize the tendency to act impulsively and engage in risky behavior when feeling positive emotions, while subtypes $\mathcal{S}2$ and $\mathcal{S}5$ tend to have lower positive urgency scores. Based on two recent studies using RS fMRIs \citep{golchert2017need, zhu2017model}, positive urgency is found to be negatively correlated with the FC within DMN in healthy subjects. It conforms to our observations that $\mathcal{S}2$ and $\mathcal{S}5$ have medium to high activation in the prime features under RS state, whereas $\mathcal{S}1$ and $\mathcal{S}6$ display low activation. In addition to the existing evidence based on RS fMRIs, our results also enhance the understanding of associations between impulsivity and FC by providing evidence from cognitive states. Our subtyping results may also provide insights on the neuronal mechanism underlying impulsivity-related disorders, such as alcohol or substance abuse and Borderline Personality Disorder \citep{cyders2008emotion,robinson2014you,fossati2014impulsivity}, which typically emerge during adolescence.

\section{Discussion} \label{Discussion}

In this paper, we propose an innovative Bayesian nonparametric clustering method for network-variates induced by multi-state brain connectivity. Leveraging the biological architecture of the brain functional system, we formulate each connectivity network-variate by stochastic block structures under each state, and simultaneously infer their community allocations as well as select the informative modular features to define each cluster. 
To facilitate the broad use of our method in real practice, we develop an efficient variational algorithm to achieve posterior computation with dramatically reduced computation and high estimation accuracy. Extensive simulations show a superior performance of our method in uncovering clusters and network architectures. By applying the model to multi-state functional connectivity data collected from children in the landmark brain cognitive development study, we establish interpretable neurodevelopment subtypes along with their brain network phenotypes.  

Currently, our method is designed to accommodate different edge types corresponding to continuous or binary networks, and it is straightforward to allow a combination of them within different states. In addition, we formulate the network-variates by stochastic block modeling here motivated by the biological architecture of functional connectivity. When different brain connectivity is considered including structural connectivity measuring the white matter anatomical connections, alternative modeling strategies on network components like latent space models could be adopted instead. With converging studies nowadays to integrate structural and functional imaging, a natural extension of our method is to perform clustering with network-variates generated from both structural and functional connectivity, where network formulation should be designed separately.

In our application to ABCD data, we focus on the fMRI data collected at baseline. Given ABCD study is designed to be longitudinal by periodically measuring different attributes of participants including neuroimaging and behaviors, it is of great interest to investigate the heterogeneity of  longitudinal brain developmental patterns to inform their future dynamics and what these could lead to. To achieve so, we could potentially incorporate a temporal domain for the networks in our method with additional models to characterize the temporal correlation between connectivity under the same state. Recent research further shows that the modular structure of the whole-brain network is also time-evolving \citep{malagurski2020longitudinal}. This indicates an interesting future direction to properly capture the temporal change of both modular structure and connectivity weights.

\bibliographystyle{biorefs}
\bibliography{refs}

\end{document}